\documentclass[a4paper,conference]{IEEEtran}
\IEEEoverridecommandlockouts

\IEEEsettopmargin{t}{30mm}
\IEEEquantizetextheight{c}
\IEEEsettextwidth{14mm}{14mm}
\IEEEsetsidemargin{c}{0mm}

\usepackage[noadjust]{cite}

\usepackage[pdftex]{graphicx}
\graphicspath{{figures/}}

\usepackage{amsmath}
\usepackage{bm}
\usepackage{amsfonts}
\usepackage{siunitx}

\DeclareSIUnit{\dB}{dB}

\usepackage{algorithm}
\usepackage[noend]{algorithmic}

\algsetup{indent=1.5em}

\usepackage{url}
\usepackage[hidelinks,pdfa]{hyperref}

\usepackage[acronym]{glossaries}

\usepackage{standalone}
\usepackage{tikz}
\usetikzlibrary{shapes.multipart, positioning, matrix, arrows.meta}
\usepackage{pgfplots}
\pgfplotsset{compat=1.17}
\usepackage{kitcolors}
\usepackage{mathtools} %
\usepackage{mleftright}

\usepackage{colorprofiles}
\usepackage[a-2b,mathxmp]{pdfx}[2018/12/22]
\hypersetup{pdfstartview=}

\pgfplotsset{
  discard if/.style 2 args={
    x filter/.append code={
      \edef\tempa{\thisrow{#1}}
      \edef\tempb{#2}
      \ifx\tempa\tempb
        
      \fi
    }
  },
  discard if not/.style 2 args={
    x filter/.append code={
      \edef\tempa{\thisrow{#1}}
      \edef\tempb{#2}
      \ifx\tempa\tempb
      \else
        
      \fi
    }
  }
}

\newacronym{ess}{ESS}{enumerative sphere shaping}
\newacronym{oess}{OESS}{optimum ESS}
\newacronym{sm}{SM}{shell mapping}
\newacronym{dm}{DM}{distribution matcher}
\newacronym{iid}{iid}{independent and identically distributed}
\newacronym{lut}{LUT}{look up table}
\newacronym{rust-package-name}{AD-ESS}{\emph{arbitrary distribution ESS}}
\newacronym{awgn}{AWGN}{additive white Gaussian noise}
\newacronym{pcs}{PCS}{probabilistic constellation shaping}
\newacronym{pas}{PAS}{probabilistic amplitude shaping}
\newacronym{fec}{FEC}{forward error correction}
\newacronym{ccdm}{CCDM}{constant composition distribution matching}
\newacronym{mpdm}{MPDM}{multiset-partition distribution matching}
\newacronym{ber}{BER}{bit error rate}
\newacronym{fer}{FER}{frame error rate}
\newacronym{ldpc}{LDPC}{low-density parity-check}
\newacronym{ask}{ASK}{amplitude-shift keying}
\newacronym{64qam}{$64$-QAM}{quadrature amplitude modulation with $64$ symbols}
\newacronym{ppc}{PPC}{peak power constrained}
\newacronym{psnr}{PSNR}{peak-signal-to-noise ratio}

\begin{document}

\title{An Extension of Enumerative Sphere Shaping for Arbitrary Channel Input Distributions}%

\author{\IEEEauthorblockN{Frederik~Ritter,
        Andrej~Rode,
        and~Laurent~Schmalen}%
        \IEEEauthorblockA{
        Communications Engineering Lab (CEL)\\
        Karlsruhe Institute of Technology (KIT)\\
        Karlsruhe, Germany\\
        email: \texttt{\{frederik.ritter, rode, schmalen\}@kit.edu}
        }%
        
}

\maketitle

\begin{abstract}
  A non-uniform channel input distribution is key for achieving the capacity of arbitrary channels.
  However, message bits are generally assumed to follow a uniform distribution which must first be transformed to a non-uniform distribution by using a distribution matching algorithm.
  One such algorithm is enumerative sphere shaping (ESS).
  Compared to algorithms such as constant composition distribution matching (CCDM), ESS can utilize more channel input symbol sequences, allowing it to achieve a comparably low rate loss.
  However, the distribution of channel input symbols produced by ESS is fixed, restricting the utility of ESS to channels with Gaussian-like capacity-achieving input distributions.
  In this paper, we generalize ESS to produce arbitrary discrete channel input distributions, making it usable on most channels.
  Crucially, our generalization replaces fixed weights used internally by ESS with weights depending on the desired channel input distribution.
  We present numerical simulations using generalized ESS with probabilistic amplitude shaping (PAS) to transmit sequences of $256$ symbols over a simplified model of an unamplified coherent optical link, a channel with a distinctly non-Gaussian capacity-achieving input distribution.
  In these simulations, we found that generalized ESS improves the maximum transmission rate by $0.0425\,\text{bit/symbol}$ at a frame error rate below $10^{-4}$ compared to CCDM.
\end{abstract}


\section{Introduction}

\IEEEPARstart{A}{pproaching} the capacity of an arbitrary channel is possible if the input symbols follow the \emph{capacity-achieving distribution} of that channel.
In the example of the \gls{awgn} channel this is a continuous normal distribution~\cite{cover_elements_2005}.
However, many communication systems are limited to independent, uniformly distributed symbols selected from a discrete set called the constellation and cannot produce the capacity-achieving distribution at the channel input.
For transmission systems limited to discrete constellations, \gls{pcs} enables the use of non-uniform channel input distributions.
This allows us to use a Maxwell-Boltzman distributed discrete input sequence, to nearly close the \textit{shaping gap} of \SI{1.53}{dB}~\cite{forney_efficient_1984,gultekin_enumerative_2020} in the \gls{awgn} channel.

By using \gls{pas}, \gls{pcs} can be combined with \gls{fec} to enable robust and flexible communications~\cite{bocherer_bandwidth_2015}.
The approximate uniform distribution of the parity bits is exploited in the \gls{pas} scheme to choose the (nearly uniformly distributed) signs of the transmitted symbols.
This works for many practically relevant channels with symmetric capacity-achieving distributions.
A \gls{dm} is responsible for the \gls{pcs} in \gls{pas}.
As the signs of the transmit symbols are defined by the parity bits, the \gls{dm} only shapes the probabilities of their amplitudes.

Multiple approaches have been proposed for implementing the \gls{dm}, e.g., \gls{ccdm}~\cite{schulte_constant_2015}.
\Gls{ccdm} maps input bit sequences to typical constant composition amplitude sequences of the desired amplitude distribution.
While \gls{ccdm} asymptotically achieves the maximum possible rate~\cite{schulte_constant_2015}, it suffers from a rate loss for finite block lengths.
\Gls{mpdm}~\cite{fehenberger_multiset-partition_2019} employs additional sequences to reduce this rate loss.
Other approaches minimize the average energy of the transmitted symbol sequences.
On the \gls{awgn} channel, this minimizes the rate loss for a fixed rate and block length~\cite{gultekin_enumerative_2020}.
Laroia's first algorithm~\cite[Alg.~1]{laroia_optimal_1994} and \gls{sm}~\cite[Alg.~2]{laroia_optimal_1994} use this technique.
This manuscript focuses on the concept of \gls{ess}~\cite{willems_pragmatic_1993}, which also minimizes the sequence energy.

The advantages of \gls{ess} in comparison with other \gls{dm} methods include a small rate loss, even at small block lengths, and low computational complexity compared to \gls{sm}~\cite{gultekin_enumerative_2020}.
One disadvantage of \gls{ess} is that it produces a fixed distribution, which is tailored to an \gls{awgn} channel.
Our contribution generalizes \gls{ess} and, based on a scheme proposed for \gls{sm}~\cite{schulte_divergence-optimal_2019}, enables the use of \gls{ess} on non-\gls{awgn} channels, while maintaining its low rate loss.

The remainder of this paper is structured as follows:
In Section~\ref{sec:generalization}, the \gls{ess} framework is generalized to use a custom weight function.
A method to create a desired weight function is introduced in Section~\ref{sec:distribution}.
Simulation results using the proposed generalized \gls{ess} are discussed in Section~\ref{sec:results}.
Finally, Section~\ref{sec:conclusion} summarizes our findings and highlights further research topics.

\section{Generalization of ESS} \label{sec:generalization}

On a high level, \gls{ess} works on all amplitude sequences with a total energy below a given threshold.
Interpreting each such sequence as a vector with amplitudes as components, all these sequences lie within a high-dimensional sphere of a radius determined by the threshold.
The sequences in this eponymous sphere are then enumerated by defining the index of each sequence as the number of lexicographically lower sequences.
This is efficiently done using a trellis representation of all amplitude sequences in the sphere.

More formally, \gls{ess} must be described in terms of this trellis.
It consists of nodes indexed by their energy $e$ and their trellis stage $n$.
Each transition between nodes is characterized by the energy difference between its source and destination node.
We call this the \emph{weight} of a transition.
In \gls{ess}, the weight of a transition is given by the square of the associated symbol amplitude, i.e., its energy.
Thus, there is a one-to-one relationship between the path through the trellis, the sequence of weights, and the sequence of symbol amplitudes.
Because the trellis does not contain nodes exceeding a fixed maximum energy, the total energy of symbol sequences represented in the trellis is also limited to this maximum energy.
Indices are assigned to all weight sequences by enumerating them in lexicographical order.
A mapping between the lexicographical index of a sequence and the data bits is established by interpreting the data bits as an integer in binary notation.
Finally, the \gls{ess} algorithm provides an efficient way to use the trellis to map between weight sequences and lexicographical indices~\cite{willems_pragmatic_1993,gultekin_enumerative_2020}.

Extensions of \gls{ess} can adapt this concept to non-energy transition weights (e.g.,~\cite{gultekin_kurtosis_2022}), thereby opening \gls{ess} to a wider range of possible output distributions.
Without changing the \gls{ess} algorithm, we can generalize the transition weights to allow for any non-negative integer.
Thus, a generalized \gls{ess} node is indexed by its trellis stage $n$ and its weight level $\ell$.
Its value is denoted as $T_n^\ell$.
Similar to \gls{ess}, the weight level of a node is the total weight of paths leading to this node.
We denote the maximum allowed weight level by $\ell_\text{max}$.

As noted in~\cite[Proposition~2]{schulte_divergence-optimal_2019} for \gls{sm}, any constant offset or positive scaling of all weights does not change the encoding, assuming $\ell_\text{max}$ is equivalently transformed.
These operations affect all the weights in the trellis equally and cannot change their order, i.e., if $w^{(1)} < w^{(2)}$ holds before scaling, $w_\mathrm{scaled}^{(1)} < w_\mathrm{scaled}^{(2)}$ will hold after scaling.
In this case, the lexicographical ordering of weight sequences remains unchanged.
Additionally, if $\ell_\text{max}$ is equivalently transformed, the set of amplitude sequences represented by the trellis does not change.
Thus, \gls{ess}, like \gls{sm}, is invariant to an offset or positive scaling of its weights.
It may be noted that the scaling factor is restricted by the aforementioned requirement of integer weights.

We consider \gls{ask} with $M$ levels and assume that the sign of each symbol is chosen using \gls{pas}.
The $M/2$ symbol amplitudes $a^{(k)} \in \{1, 3, 5, \dots, M-1\}$, $k \in \mathcal{K} = \{0, 1, \dots, \frac{M}{2}-1\}$ are chosen using our proposed generalization of \gls{ess}, that is, by using a trellis with transition weights not fixed to the amplitude energy.
Instead, each amplitude $a^{(k)}$ is associated with a general weight denoted~$w^{(k)}$.
Without loss of generality, {\small $\mleft(w^{(k)}\mright)_{k \in \mathcal{K}}$} is assumed to be in ascending order, i.e., {\small $w^{(k_1)} \leq w^{(k_2)}$ if ${k_1 < k_2}$}.
Note that this implies that $\left(a^{(k)}\right)_{k \in \mathcal{K}}$ is not necessarily ordered but depends on the amplitude to weight mapping.
More specifically, the amplitudes are not ordered, if the correspondence between amplitudes and weights is chosen in such a way that $a^{(k_1)} < a^{(k_2)}$ does not imply $w^{(k_1)} < w^{(k_2)}$.
As the \gls{ess} trellis is invariant to a constant offset, we require
\begin{equation}
  \underset {k \in \mathcal{K}} \min \, w^{(k)} = w^{(0)} = 0,
  \label{eq:min-weight-req}
\end{equation}
which allows the set of weight levels $\mathcal{L} \subset \{0, 1, \dots, \ell_\text{max}\}$ to be independent of the trellis stage $n$.

Verification that this framework does indeed generalize \gls{ess} can be obtained by using the amplitude energies $(1, \, 9, \, 25, \, 49)$ as weights.
By~\eqref{eq:min-weight-req}, the value $1$ is subtracted from all weights which leads to the valid weight sequence $(0, \, 8, \, 24, \, 48)$.
It is a good practice to use the smallest possible scaling of the weight function, therefore all weights are divided by $8$ which yields the final weights $(0, \, 1, \, 3, \, 6)$.
A trellis with these weights is identical to a classical \gls{ess} trellis.
The only remaining differences are the node indices which can easily be converted from weight level $\ell$ to energy $e$ via $e = 8\ell + n$.
This weight function was also proposed in~\cite{jiang_non-recursive_2023} as a more efficient method to calculate the \gls{ess} trellis. %

\begin{figure}
  \begin{center}
  \input{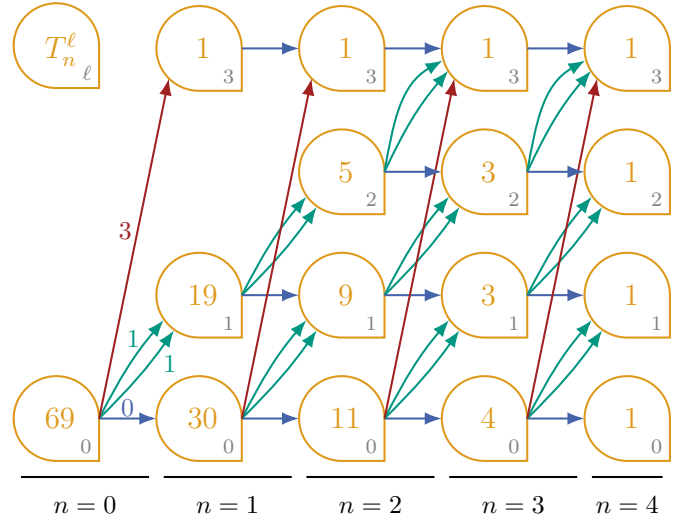}%
\newcount\b%
\begin{tikzpicture}[thick]
\tikzset{defaultnode/.style={draw,trellisnode,color=KITorange,inner sep=3pt, text width=2ex, align=right}}%
\matrix (trellis)
[matrix of nodes, nodes={trellisnode, draw}, column sep={0.75cm},
row sep={0.5cm}]
{%
  \node[trellisnode, defaultnode](trellis-1-1){\nodepart{upper}{\large $T^\ell_n$}\nodepart[color=gray]{lower}\footnotesize$\ell$};
&\node[trellisnode, defaultnode](trellis-1-2){\nodepart{upper}{\large$1$}\nodepart[color=gray]{lower}\footnotesize$3$};
&\node[trellisnode, defaultnode](trellis-1-3){\nodepart{upper}{\large$1$}\nodepart[color=gray]{lower}\footnotesize$3$};
&\node[trellisnode, defaultnode](trellis-1-4){\nodepart{upper}{\large$1$}\nodepart[color=gray]{lower}\footnotesize$3$};
&\node[trellisnode, defaultnode](trellis-1-5){\nodepart{upper}{\large$1$}\nodepart[color=gray]{lower}\footnotesize$3$};\\
\node[coordinate](trellis-2-1){};
& \node[coordinate](trellis-2-2){};
&\node[trellisnode, defaultnode](trellis-2-3){\nodepart{upper}\large$5$\nodepart[color=gray]{lower}\footnotesize$2$};
&\node[trellisnode, defaultnode](trellis-2-4){\nodepart{upper}\large$3$\nodepart[color=gray]{lower}\footnotesize$2$};
&\node[trellisnode, defaultnode](trellis-2-5){\nodepart{upper}\large$1$\nodepart[color=gray]{lower}\footnotesize$2$};\\
\node[coordinate](trellis-3-1){};
&\node[trellisnode, defaultnode](trellis-3-2){\nodepart{upper}\large$19$\nodepart[color=gray]{lower}\footnotesize$1$};
&\node[trellisnode, defaultnode](trellis-3-3){\nodepart{upper}\large$9$\nodepart[color=gray]{lower}\footnotesize$1$};
&\node[trellisnode, defaultnode](trellis-3-4){\nodepart{upper}\large$3$\nodepart[color=gray]{lower}\footnotesize$1$};
&\node[trellisnode, defaultnode](trellis-3-5){\nodepart{upper}\large$1$\nodepart[color=gray]{lower}\footnotesize$1$};\\
\node[trellisnode, defaultnode](trellis-4-1){\nodepart{upper}\large$69$\nodepart[color=gray]{lower}\footnotesize$0$};
&\node[trellisnode, defaultnode](trellis-4-2){\nodepart{upper}\large$30$\nodepart[color=gray]{lower}\footnotesize$0$};
&\node[trellisnode, defaultnode](trellis-4-3){\nodepart{upper}\large$11$\nodepart[color=gray]{lower}\footnotesize$0$};
&\node[trellisnode, defaultnode](trellis-4-4){\nodepart{upper}\large$4$\nodepart[color=gray]{lower}\footnotesize$0$};
&\node[trellisnode, defaultnode](trellis-4-5){\nodepart{upper}\large$1$\nodepart[color=gray]{lower}\footnotesize$0$};\\
};%
\foreach[count=\a] \b in {2,...,5}{%

    \ifnum\a>1{
    \draw [KITblue, -{Latex}] (trellis-4-\a.east) -- (trellis-4-\b.west);%
    \draw [KITgreen] (trellis-4-\a.east) edge[out=45,in=235,-{Latex}] ([xshift=2,yshift=-2]trellis-3-\b.south west);
    \draw [KITgreen] (trellis-4-\a.east) edge[out=63,in=230,-{Latex}] ([xshift=-2,yshift=2]trellis-3-\b.south west);
    \draw [KITgreen] (trellis-3-\a.east) edge[out=45,in=235,-{Latex}] ([xshift=2,yshift=-2]trellis-2-\b.south west);
    \draw [KITgreen] (trellis-3-\a.east) edge[out=63,in=230,-{Latex}] ([xshift=-2,yshift=2]trellis-2-\b.south west);
    \draw [KITblue, -{Latex}] (trellis-3-\a.east) -- (trellis-3-\b.west);%
    \draw [KITblue, -{Latex}] (trellis-1-\a.east) -- (trellis-1-\b.west);%
    \draw [KITred, -{Latex}] (trellis-4-\a.east) -- (trellis-1-\b.south west);%
    \ifnum\a>2{%
    \draw [KITgreen] (trellis-2-\a.east) edge[out=80,in=210,-{Latex}] ([xshift=-4.5,yshift=7]trellis-1-\b.south west);
    \draw [KITgreen] (trellis-2-\a.east) edge[out=65,in=230,-{Latex}] ([xshift=-2.3,yshift=3]trellis-1-\b.south west);
    \draw [KITblue, -{Latex}] (trellis-2-\a.east) -- (trellis-2-\b.west);%
    }\fi%
    }%
    \else{%
    \draw [KITblue, -{Latex}] (trellis-4-\a.east) -- node[above, yshift=-0.1cm]{$0$} (trellis-4-\b.west);%
    \draw [KITgreen] (trellis-4-\a.east) edge[out=45,in=235,-{Latex}] node[midway, below, xshift=12, yshift=12]{$1$} ([xshift=2,yshift=-2]trellis-3-\b.south west);
    \draw [KITgreen] (trellis-4-\a.east) edge[out=63,in=230,-{Latex}] node[midway, above, xshift=2, yshift=4]{$1$} ([xshift=-2,yshift=2]trellis-3-\b.south west);
    \draw [KITred, -{Latex}] (trellis-4-\a.east) -- node[midway, above left, xshift=0.1cm]{$3$} (trellis-1-\b.south west);%
    }%
    \fi%
    \draw [black] ([yshift=-0.8cm,xshift=0.1cm]trellis-4-\a.west) -- node[midway](line-\a){} ([yshift=-0.8cm,xshift=-0.1cm]trellis-4-\b.west);%
}%
\node[rectangle, below= 0.0 of line-1]{$n=0$};%
\node[rectangle, below= 0.0 of line-2]{$n=1$};%
\node[rectangle, below= 0.0 of line-3]{$n=2$};%
\node[rectangle, below= 0.0 of line-4]{$n=3$};%
\draw [black] ([yshift=-0.8cm,xshift=0.1cm]trellis-4-5.west) -- node[midway](line-5){} ([yshift=-0.8cm,xshift=-0.1cm]trellis-4-5.east);%
\node[rectangle, below= 0.0 of line-5]{$n=4$};%
\end{tikzpicture}
  \end{center}
  \caption{Generalized \gls{ess} trellis with $N=4$ and non-unique weights $(0, \, 1, \, 1, \, 3)$.}
  \label{fig:non_unique_trellis}
\end{figure}

\subsection{Non-unique weights} %

Considering only the classical \gls{ess} trellis, our proposed generalization is subject to the additional restriction of unique weights.
However, multiple identical weights can be modelled by allowing parallel edges in the trellis, as shown in Fig.~\ref{fig:non_unique_trellis}.
The \gls{ess} algorithm must then be adapted to enforce an order on these parallel edges.
We propose ordering parallel edges based on the index $k$ of their associated weight $w^{(k)}$.
Algorithm~\ref{alg:encode} shows the \gls{ess} shaping algorithm as formulated in~\cite[Alg.~1]{gultekin_enumerative_2020}, modified for generalized \gls{ess} with parallel edges.
Comparison with~\cite[Alg.~1]{gultekin_enumerative_2020} shows two modifications:
Primarily, the use of the index variable $k$ instead of the amplitude $a$, which allows for parallel edges in the trellis.
Secondarily, the use of zero-based indexing, which ensures consistency with the notation used in this paper.
Similarly, Algorithm~\ref{alg:decode} adapts \gls{ess} deshaping from~\cite[Alg.~2]{gultekin_enumerative_2020} for generalized \gls{ess} with non-unique weights.
The similarities between~\cite{gultekin_enumerative_2020} and the two algorithms shown here highlight the simplicity of handling non-unique weights with the proposed index-based approach.

\begin{algorithm}[tb]
  \caption{Generalized Enumerative Shaping}
  \label{alg:encode}
  Given that the index satisfies $0 \leq i < T_0^0$,
  initialize the algorithm by setting the \textit{local index} $i_0 = i$.
  Then for ${n = 0,1,\dots,N-1}$:
  \begin{enumerate}
    \item
      Take $k_n \in \mathcal{K}$ such that
  \end{enumerate}
      \begin{equation}
        \sum_{k' < k_n} T_n^{w^{(k')} + \sum_{j=0}^{n-1} w^{(k_j)}} \leq i_n < \sum_{k' \leq k_n} T_n^{w^{(k')} + \sum_{j=0}^{n-1} w^{(k_j)}} \text{,}
      \end{equation}
  \begin{enumerate}
    \setcounter{enumi}{1}
    \item
      and (for $n < N$)
      \begin{equation}
        i_{n+1} = i_n - \sum_{k'<k_n} T_n^{w^{(k')} + \sum_{j=0}^{n-1} w^{(k_j)}} \text{.}
      \end{equation}
  \end{enumerate}
  Finally output $a^{(k_0)}, a^{(k_1)}, \dots, a^{(k_{N-1})}$.
\end{algorithm}

\begin{algorithm}[tb]
  \caption{Generalized Enumerative Deshaping}
  \label{alg:decode}
  Given $a_0, a_1, \dots, a_{N-1}$:
  Derive $k_0, k_1, \dots, k_{N-1}$ s.t. $a_n = a^{(k_n)}$
  \begin{enumerate}
    \item
      Initialize the algorithm by setting the \textit{local index} $i_N = 0$
    \item
      For $k' \in \mathcal{K}$ and $n = N-1, N-2, \dots, 0$, update the local index as
      \begin{equation}
        i_{n} = \sum_{k'<k_n} T_n^{w^{(k')} + \sum_{j=0}^{n-1} w^{(k_j)}} + i_{n+1} \text{.}
      \end{equation}
    \item
      Finally output $i = i_0$.
  \end{enumerate}
\end{algorithm}

\section{Choosing a Distribution via Weights} \label{sec:distribution}

\subsection{Divergence-Optimal Weights}

Generalized \gls{ess} opens the \gls{ess} algorithm to a wide range of distributions.
This calls for a method to choose the amplitude weights in such a way that the resulting distribution approaches the capacity-achieving input distribution for a given channel.
In~\cite{schulte_divergence-optimal_2019}, Schulte and Steiner develop such a method for \gls{sm}.
Both \gls{sm} and \gls{ess} work according to the same principle of generating a code book which minimizes the weight of its codewords.
Thus, the approach proposed for \gls{sm} can also be used for \gls{ess}.
We summarize the method developed in~\cite{schulte_divergence-optimal_2019} in the context of \gls{ess}.

First, the informational divergence $\mathbb{D}(U_{\bm{A}}||P_{\bm{A}})$ is introduced.
The distribution $P_{\bm{A}}$ is the capacity-achieving input distribution for the channel in question, and $U_{\bm{A}}(\bm{a})$ is the probability of transmitting the amplitude sequence ${\bm{a} = (a_0, \, a_1, \, \dots, \, a_{N-1})}$.
As the \gls{dm} chooses uniformly from the $|\mathcal{C}|$ amplitude sequences in its code book $\mathcal{C}$, ${U_{\bm{A}}(\bm{a}) = |\mathcal{C}|^{-1}}$ follows for all amplitude sequences $\bm{a} \in \mathcal{C}$.
The informational divergence $\mathbb{D}(U_{\bm{A}}||P_{\bm{A}})$ therefore depends on the capacity-achieving distribution and the selection of amplitude sequences in the \gls{dm} code book.
It can be shown, that the mutual information $\mathbb{I}(\bm{A};\bm{Y})$ between an input amplitude sequence $\bm{A}$ and the channel output $\bm{Y}$ is bounded by~\cite[Eq.~(7)]{schulte_divergence-optimal_2019},~\cite[Eq.~(23)]{bocherer_matching_2011}
\begin{equation}
  C - \frac{\mathbb{D}(U_{\bm{A}}||P_{\bm{A}})}{N} \leq \frac{\mathbb{I}(\bm{A};\bm{Y})}{N} \leq C.
  \label{eq:mi-bound}
\end{equation}

Minimizing the informational divergence $\mathbb{D}(U_{\bm{A}}||P_{\bm{A}})$ thus bounds the mutual information closer to the channel capacity $C$.
Assuming that the input amplitudes are \gls{iid} according to the capacity-achieving distribution, the probability $P_{\bm{A}}$ can be written as
\begin{equation*}
  P_{\bm{A}}(\bm{a}) = \prod_{i=0}^{N-1} P_A(a_i).
\end{equation*}

If each weight of generalized \gls{ess} is defined as the self-information $-\log(P(a))$ of the corresponding amplitude $a$, the resulting code book minimizes $\mathbb{D}(U_{\bm{A}}||P_{\bm{A}})$~\cite[Proposition~1]{schulte_divergence-optimal_2019}.
This can be verified by expanding the expression of the informational divergence
\begin{align}
  \mathbb{D}(U_{\bm{A}}||P_{\bm{A}})
    &= \sum_{\bm{a} \in \mathcal{C}} U_{\bm{A}}(\bm{a}) \cdot \log \frac{U_{\bm{A}}(\bm{a})}{P_{\bm{A}}(\bm{a})} \nonumber \\
    &= - \mathbb{H}_U(\bm{A}) - \sum_{\bm{a} \in \mathcal{C}} U_{\bm{A}}(\bm{a}) \cdot \log P_{\bm{A}}(\bm{a}) \nonumber \\
    &= - \log |\mathcal{C}| + \frac{1}{|\mathcal{C}|} \cdot \sum_{\bm{a} \in \mathcal{C}} \sum_{n=0}^{N-1} \left(- \log P_A(a_n)\right), \label{eq:divergence-expanded}
\end{align}
where we expand the logarithm of the fraction in the first step and recognize the entropy.

With $w^{(k)} \coloneq - \log P_A(a^{(k)})$ defined as the weight of amplitude $a^{(k)}$, the double sum in~\eqref{eq:divergence-expanded} becomes the total weight of the code book
\begin{align*}
  W(\mathcal{C}) &\coloneq \sum_{\bm{a} \in \mathcal{C}} \sum_{n=0}^{N-1} w^{(k_n)} \quad \text{with} \quad a_n = a^{\left(k_n\right)}, \\
  &= \sum_{\bm{a} \in \mathcal{C}} \sum_{n=0}^{N-1} \left(- \log P_A(a_n)\right).
\end{align*}
For a given $|\mathcal{C}|$ and a given set of discrete amplitudes/weights, generalized \gls{ess} minimizes the total weight~$W(\mathcal{C})$ of the code book.
It therefore minimizes $\mathbb{D}(U_{\bm{A}}||P_{\bm{A}})$ for a fixed~$|\mathcal{C}|$.
As a result, generalized \gls{ess} maximizes the lower bound~\eqref{eq:mi-bound} on the mutual information.
In practice, this is often only approximately true due to the requirement for weights to be integer and the code book size to be a power of two.

Note, that the informational divergence between the distribution \textit{of the sequences} in $\mathcal{C}$ and the capacity-achieving distribution \textit{of sequences} is minimized.
Somewhat counter intuitively, the informational divergence between the empirical \textit{amplitude} distribution and the capacity-achieving \textit{amplitude} distribution is not minimized.
Equation~(12) in~\cite{schulte_divergence-optimal_2019} details the dependencies between the informational divergences of the sequence and amplitude distributions.

\subsection{Implementation Aspects}

For implementation, two issues which have not yet been discussed, arise from this approach. %
First, the self-information of the amplitudes is not integer in general.
Second, the calculation of all possible weight levels becomes necessary to store the trellis.
We present possible solutions for both issues in what follows.

The requirement of integer weights forces the quantization of the self-information weight function.
As previously discussed, positive scaling and a constant offset applied to the weights do not change the values in the trellis.
Scaling the self-information with a factor $f > 1$ before quantizing can thus reduce the relative quantization error.
Assuming sorted weights and additionally respecting the requirement~\eqref{eq:min-weight-req} for the minimum weight to be zero, we propose the positive integer weight function:

\begin{equation}
  \begin{gathered} \label{eq:weight-function}
    w^{(k)} = \omega^{(k)} - \omega^{(0)} \\
    \text{with} \quad \omega^{(k)} = \left\lfloor - f \cdot \log P_A\left(a^{(k)}\right) + \frac{1}{2} \right\rfloor.
  \end{gathered}
\end{equation}

Choosing $f$ is a trade-off between low quantization noise and trellis size.
If $f$ is too large, the probability that sums of different weights lead to the same weight level decreases.
For instance, the distribution $(0.4, 0.3, 0.2, 0.1)$ leads to the weights $(0, 1, 2, 4)$ with $f = 3$, but to $(0, 3, 7, 14)$ with $f = 10$.
One can easily verify that, e.g., $w^{(1)} + w^{(1)} = w^{(2)}$ is true for $f = 3$ but is not for $f = 10$.
Thus $f = 10$ has an additional weight level $\ell_2 = w^{(1)} + w^{(1)} = 6$ between $\ell_1 = w^{(1)} = 3$ and $\ell_3 = w^{(2)} = 7$.
This increases the size of the trellis.
On a more general note, the same effect would balloon the trellis size if non-integer weights were used.

In classical \gls{ess}, the index of a node in its trellis stage can easily be computed from its energy $e$ and trellis stage number~$n$~\cite[Sec.~III.~B]{gultekin_enumerative_2020}. %
With generalized weights, this is no longer universally possible.
A \gls{lut} between weight levels $\ell$ and the corresponding trellis row indices must thus be stored if the trellis rows are kept in an array.
Considering that $\log_2(|\mathcal{L}|)$ bit are required to store one of the $|\mathcal{L}|$ trellis row indices and $\log_2(\ell_\text{max})$ bit are required to store a weight level, the \gls{lut} storage complexity is $(\log_2(|\mathcal{L}|) + \log_2(\ell_\text{max})) \cdot |\mathcal{L}|$.
In analogy to \gls{ess}, the number $|\mathcal{L}|$ of nodes in a trellis stage is expected to be roughly proportional to the amplitude sequence length~$N$~\cite[Sec.~IV.~B]{gultekin_enumerative_2020}.
Additionally, assuming $\ell_\text{max}$ is proportional to $|\mathcal{L}|$, we can express the approximate storage complexity of the \gls{lut} as $N \, \log_2(N)$.
Compared to the storage complexity of the trellis itself, which is approximately proportional to~$N^3$~\cite[Tab.~II]{gultekin_enumerative_2020}, the \gls{lut} thus does not constitute a significant extra complexity.

With the use of a \gls{lut}, the calculation of weight levels only needs to be carried out once.
Thus, no stringent complexity limits must be considered for this calculation.
A simple algorithm which iteratively creates new weight levels by addition of known weight levels proved sufficient in our experience.
It is summarized in Algorithm~\ref{alg:weight-levels}.

\begin{algorithm}[t]
  \caption{Calculation of Weight Levels}
  \label{alg:weight-levels}
  \begin{algorithmic}
    \REQUIRE Weights $w^{(0)},w^{(1)},\dots,w^{(M-1)}$
    \STATE $\mathcal{L} \coloneq \{0\}$
    \REPEAT
      \STATE $\mathcal{L}' \coloneq \mathcal{L}$
      \FOR{$\ell \in \mathcal{L}'$}
        \FOR{$i \in \{0, 1, \dots, M-1\}$}
          \STATE $\ell_\text{new} \coloneq \ell + w^{(i)}$
          \IF{$\ell_\text{new} \leq \ell_\text{max}$}
            \STATE $\mathcal{L} \coloneq \{\ell_\text{new}\} \cup \mathcal{L}$
          \ENDIF
        \ENDFOR
      \ENDFOR
    \UNTIL{$|\mathcal{L}| = |\mathcal{L}'|$}
    \RETURN $\mathcal{L}$
  \end{algorithmic}
\end{algorithm}

Generalized \gls{ess} would benefit from further research into a more efficient handling of the irregular weight levels.
One approach may be to choose $f$ in~\eqref{eq:weight-function} in such a way that ${w^{(1)} = 1}$.
In this case, the weights would collapse to relative indices in the arrays used to store the trellis values, removing the necessity for a \gls{lut}.
However, further investigation is required to weigh the resulting potentially coarse quantisation of weight levels against the reduced complexity.

\subsection{Open Source Implementation}

We provide a Rust implementation of the discussed \gls{ess} generalization called \gls{rust-package-name}\footnote{Free source code at \texttt{https://github.com/kit-cel/ad-ess}}.
Alongside the Rust implementation, we also publish Python bindings, which allow using the Rust binaries from Python scripts.

\section{Simulation Results} \label{sec:results}

One advantage of \gls{pcs} using the \gls{pas} architecture is the ability to adapt the transmission rate using only a small number of \gls{fec} code rates.
This is done by changing the shaping rate of the \gls{dm}, which corresponds to the number of bits the \gls{dm} maps to one amplitude sequence.
To demonstrate the proposed generalization of \gls{ess}, we selected four WiMAX \gls{ldpc} codes.
The four codes have a block length of $768$ bits and rates of $r_\text{LDPC} \in \{1/2, 2/3, 3/4, 5/6\}$.
An $M=8$-\gls{ask} is used as modulation format to simulate one dimension of a \gls{64qam}.
Following the \gls{pas} architecture, the fixed \gls{ldpc} block length and constellation size lead to a fixed sequence length of $256$ symbols.

These sequences are transmitted over a \gls{ppc} channel with \gls{awgn}, which is a coarse model of an unamplified coherent optical link~\cite{che_does_2021}.
To quantify the channel quality, the \gls{psnr} is defined as the maximum signal power divided by the noise power (similar to~\cite{che_does_2021}).
As only the maximum signal power affects the \gls{psnr}, it is beneficial to use the full range of power below the maximum to increase the spacing between transmitted amplitudes.
In contrast, the Maxwell-Boltzmann distribution assigns high probabilities to small amplitudes, which are all relatively close together.
This makes the Maxwell-Boltzmann distribution ill-suited for the \gls{ppc} channel.

One approach is a reversed Maxwell-Boltzmann distribution which assigns high probability to high-amplitude symbols and low probability to low-amplitude symbols~\cite{che_does_2021}.
We implemented this using a two-step process we call reverse \gls{ess}.
In the first step, the data is encoded using unaltered \gls{ess} resulting in a sequence of approximately Maxwell-Boltzmann distributed amplitudes.
Then, in the second step, each amplitude in this sequence is remapped according to $a \mapsto M - a$, which effectively swaps low and high amplitude values.
The resulting amplitude sequence is approximately distributed according to a reversed Maxwell-Boltzmann distribution.

To the best of the authors' knowledge, there is no closed-form solution for the capacity-achieving input distribution of the \gls{ppc} \gls{awgn} channel with $8$-\gls{ask} input.
Hence, following the example of \cite{oliveira_capacity-achieving_2023}, we approximate it using numerical optimization for each \gls{psnr} value.
We formulate this optimization problem with quantized channel outputs and solve it using CVXPY~\cite{diamond_cvxpy_2016, agrawal_rewriting_2018}.
Channel inputs following the resulting distributions can be implemented with generalized ESS.

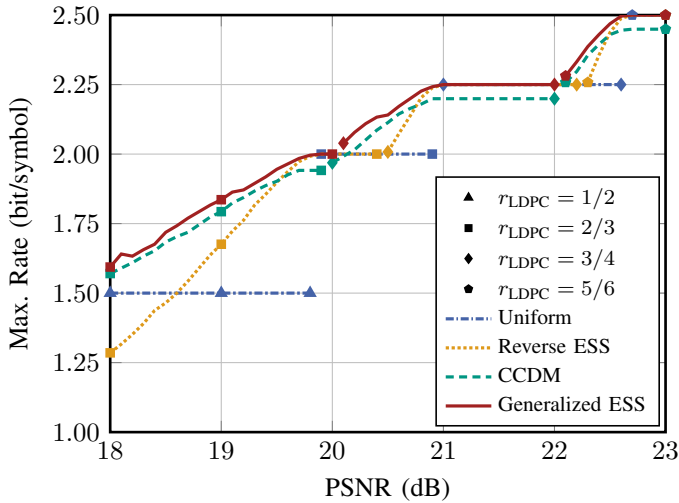
\begin{figure}
  \begin{tikzpicture}
    \tikzset{
      cfg1/.style={mark=triangle*, mark options={solid}, mark size=1.7pt},
      cfg2/.style={mark=square*, mark options={solid}, mark size=1.2pt},
      cfg3/.style={mark=diamond*, mark options={solid}, mark size=1.7pt},
      cfg4/.style={mark=pentagon*, mark options={solid}, mark size=1.5pt},
      adts/.style={KITred, solid, line width=1.2pt},
      rev_ess/.style={KITorange, densely dotted, line width=1.2pt},
      ccdm/.style={KITgreen, densely dashed, line width=1.2pt},
      no_dm/.style={KITblue, densely dashdotted, line width=1.2pt},
    }
    \begin{axis}[
        xlabel={PSNR (dB)},
        ylabel={Max. Rate (bit/symbol)},
        ymin=1,
        ymax=2.5,
        ytick={1, 1.25, 1.5, 1.75, 2, 2.25, 2.5, 2.75},
        xmin=18.0,
        xmax=23.0,
        grid=both,
        width=1\columnwidth,
        height=0.8\columnwidth,
        clip mode=individual,
        line width=1pt,
        legend style={
          at={(axis cs:22.95,1.02)},
          anchor=south east,
          nodes={transform shape},
          font=\footnotesize,
          line width=0.6pt
        },
        legend cell align={left},
        y tick label style={
          /pgf/number format/.cd,
          fixed,
          fixed zerofill,
          precision=2,
          /tikz/.cd
        },%
        x tick label style={
          /pgf/number format/.cd,
          fixed,
          /tikz/.cd
        },%
      ]%

      \addplot+[%
        forget plot,
        mark indices={1,11,19},
        cfg1, no_dm%
        ]%
        table[discard if not={heuristic_NO_DM_config}{n768_r0.5_m3}, x=psnr_dB, y=heuristic_NO_DM_rate, col sep=comma,
        ]{./data/optimum_rates_andrej_fine_FER-0.0001_CH-AWGN_PPC.csv};%
      \addplot+[%
        forget plot,
        mark indices={1,11},
        cfg2, no_dm%
        ]%
        table[discard if not={heuristic_NO_DM_config}{n768_r0.67_m3}, x=psnr_dB, y=heuristic_NO_DM_rate, col sep=comma,
        ]{./data/optimum_rates_andrej_fine_FER-0.0001_CH-AWGN_PPC.csv};%
      \addplot+[%
        forget plot,
        mark indices={1,17},
        cfg3, no_dm%
        ]%
        table[discard if not={heuristic_NO_DM_config}{n768_r0.75_m3}, x=psnr_dB, y=heuristic_NO_DM_rate, col sep=comma,
        ]{./data/optimum_rates_andrej_fine_FER-0.0001_CH-AWGN_PPC.csv};%
      \addplot+[%
        forget plot,
        mark indices={1,4},
        cfg4, no_dm%
        ]%
        table[discard if not={heuristic_NO_DM_config}{n768_r0.83_m3}, x=psnr_dB, y=heuristic_NO_DM_rate, col sep=comma,
        ]{./data/optimum_rates_andrej_fine_FER-0.0001_CH-AWGN_PPC.csv};%

      \addplot+[%
        forget plot,
        mark indices={1,11,25},
        cfg2, rev_ess%
        ]%
        table[discard if not={heuristic_Rev_ESS_config}{n768_r0.67_m3}, x=psnr_dB, y=heuristic_Rev_ESS_rate, col sep=comma,
        ]{./data/optimum_rates_andrej_fine_FER-0.0001_CH-AWGN_PPC.csv};%
      \addplot+[%
        forget plot,
        mark indices={1,18},
        cfg3, rev_ess%
        ]%
        table[discard if not={heuristic_Rev_ESS_config}{n768_r0.75_m3}, x=psnr_dB, y=heuristic_Rev_ESS_rate, col sep=comma,
        ]{./data/optimum_rates_andrej_fine_FER-0.0001_CH-AWGN_PPC.csv};%
      \addplot+[%
        forget plot,
        mark indices={1,8},
        cfg4, rev_ess%
        ]%
        table[discard if not={heuristic_Rev_ESS_config}{n768_r0.83_m3}, x=psnr_dB, y=heuristic_Rev_ESS_rate, col sep=comma,
        ]{./data/optimum_rates_andrej_fine_FER-0.0001_CH-AWGN_PPC.csv};%
      \addplot+[%
        forget plot,
        mark indices={1,11,20},
        cfg2, ccdm%
        ]%
        table[discard if not={heuristic_CCDM_config}{n768_r0.67_m3}, x=psnr_dB, y=heuristic_CCDM_rate, col sep=comma,
        ]{./data/optimum_rates_andrej_fine_FER-0.0001_CH-AWGN_PPC.csv};%
      \addplot+[%
        forget plot,
        mark indices={1,21},
        cfg3, ccdm%
        ]%
        table[discard if not={heuristic_CCDM_config}{n768_r0.75_m3}, x=psnr_dB, y=heuristic_CCDM_rate, col sep=comma,
        ]{./data/optimum_rates_andrej_fine_FER-0.0001_CH-AWGN_PPC.csv};%
      \addplot+[%
        forget plot,
        mark indices={1,10},
        cfg4, ccdm%
        ]%
        table[discard if not={heuristic_CCDM_config}{n768_r0.83_m3}, x=psnr_dB, y=heuristic_CCDM_rate, col sep=comma,
        ]{./data/optimum_rates_andrej_fine_FER-0.0001_CH-AWGN_PPC.csv};%

      \addplot+[%
        forget plot,
        mark indices={1,11,21},
        cfg2, adts%
        ]%
        table[discard if not={heuristic_ADTS_config}{n768_r0.67_m3}, x=psnr_dB, y=heuristic_ADTS_rate, col sep=comma,
        ]{./data/optimum_rates_andrej_fine_FER-0.0001_CH-AWGN_PPC.csv};%
      \addplot+[%
        forget plot,
        mark indices={1,20},
        cfg3, adts%
        ]%
        table[discard if not={heuristic_ADTS_config}{n768_r0.75_m3}, x=psnr_dB, y=heuristic_ADTS_rate, col sep=comma,
        ]{./data/optimum_rates_andrej_fine_FER-0.0001_CH-AWGN_PPC.csv};%
      \addplot+[%
        forget plot,
        mark indices={1,10},
        cfg4, adts%
        ]%
        table[discard if not={heuristic_ADTS_config}{n768_r0.83_m3}, x=psnr_dB, y=heuristic_ADTS_rate, col sep=comma,
        ]{./data/optimum_rates_andrej_fine_FER-0.0001_CH-AWGN_PPC.csv};%

      \addplot+[cfg1, black, only marks] coordinates{(0, 0)};%
      \addplot+[cfg2, black, only marks] coordinates{(0, 0)};%
      \addplot+[cfg3, black, only marks] coordinates{(0, 0)};%
      \addplot+[cfg4, black, only marks] coordinates{(0, 0)};%
      \addplot+[no_dm, mark=none] coordinates{(0, 0)};%
      \addplot+[rev_ess, mark=none] coordinates{(0, 0)};%
      \addplot+[ccdm, mark=none] coordinates{(0, 0)};%
      \addplot+[adts, mark=none] coordinates{(0, 0)};%

      \legend{%
        $r_\text{LDPC} = 1/2$,%
        $r_\text{LDPC} = 2/3$,%
        $r_\text{LDPC} = 3/4$,%
        $r_\text{LDPC} = 5/6$,%
        Uniform,
        Reverse ESS,
        CCDM,
        Generalized ESS,
        Capacity
      };
    \end{axis}%
  \end{tikzpicture}%
  \caption{Maximum rates achieving a \gls{fer} below $10^{-4}$ for a fixed sequence length of $256$ $8$-\gls{ask} symbols.}%
  \label{fig::max-rate-plot-fer}%
\end{figure}

The largest achievable rate using any of the four available \gls{ldpc} codes and an \gls{fer} below $10^{-4}$ is shown in Fig.~\ref{fig::max-rate-plot-fer}.
Generalized \gls{ess} is compared to \gls{ccdm} and the reverse Maxwell-Boltzmann approach implemented as reversed \gls{ess}.
Rates achievable using the four codes with uniform signaling are included for reference.

Generalized \gls{ess} and reversed \gls{ess} consistently match or improve the rate of uniform signaling.
The reason for this is that both of these trellis-based approaches are true generalizations of uniform signaling:
If the maximum weight level $\ell_\text{max}$ or energy threshold is increased until all possible sequences are represented in the trellis, the \gls{dm} outputs a uniform distribution over all possible amplitude sequences.
This is not true for \gls{ccdm}, which only outputs sequences of constant composition, thus, resulting in a rate loss at finite block lengths.

While the two trellis-based approaches generalize uniform signaling, this is not without limitations.
First, the use of \gls{pas} with an $8$-\gls{ask} only allows code rates $r_\text{LDPC} \geq 2/3$.
This explains why reversed \gls{ess} cannot match the rate of uniform signaling with $r_\text{LDPC} = 1/2$ between $18$ and \qty{18.6}{\decibel}.
The constellation order has to be reduced to use lower rate codes.
Second, we observed that erroneously received frames with shaping contain more bit errors than erroneous frames in systems without shaping.
Finally, using a trellis for uniform signaling is computationally inefficient as it requires computing and storing the full trellis without limiting it to a maximum weight level~$\ell_\text{max}$ or an energy threshold.

In an operating point where the maximum rate supported by the channel lies between the rates of two available codes, uniform signaling must use the lower rate code.
Here, the use of a \gls{dm} introduces the shaping rate as an additional variable that can be adjusted to better match the maximum rate supported by the channel.
Fig.~\ref{fig::max-rate-plot-fer} shows this behavior in the intervals $18$~to~\qty{20}{\decibel}, $20$~to~\qty{21}{\decibel} and $22$~to~\qty{22.5}{\decibel}.
By using an amplitude distribution optimized for the channel, generalized \gls{ess} can achieve the highest rates of all simulated schemes in these intervals.
Comparing generalized \gls{ess} to \gls{ccdm} shows that the two rate curves are almost parallel, as \gls{ccdm} uses the same optimized distribution but suffers from a rate loss at finite lengths.
On average, the rate loss of \gls{ccdm} compared to generalized \gls{ess} is $0.0425\,\text{bit/symbol}$.
The amplitude distribution used by reversed \gls{ess} is less suited for the channel.
Thus, shaping would lead to lower rates than uniform signaling with the next lower code rate for many \gls{psnr} values.
For these \gls{psnr} values, the highest rates are achieved with reversed \gls{ess} by resorting to uniform signaling with the next lower code rate.
Fig.~\ref{fig::max-rate-plot-fer} shows this behaviour for \gls{psnr} values between $19.9$ and \qty{20.4}{\decibel} or $21$ and \qty{22.2}{\decibel}.

To change the shaping rate of \gls{ccdm} and thus enable rate adaption, target distributions with different entropies must be used.
A method to find such distributions is proposed in~\cite{bocherer_bandwidth_2015}.
The proposed method relies on Maxwell-Boltzmann distributions and can not be used on our channel.
However, inspired by this method we use the heuristic
\begin{equation*}
  P_\text{mod.}(a^{(k)}) = \frac
      {P_\text{opt.}(a^{(k)}) \cdot (P_\text{opt.}(a^{(k)}))^\lambda}
      {\sum_{k \in \mathcal{K}} P_\text{opt.}(a^{(k)}) \cdot (P_\text{opt.}(a^{(k)}))^\lambda}
\end{equation*}
to generate distributions with suitable entropy from the optimized distribution by tuning $\lambda$.
As a result, the \gls{ccdm} rates shown in Fig.~\ref{fig::max-rate-plot-fer} are not guaranteed to show the best possible \gls{ccdm} performance.
Generalized \gls{ess} has no need for such a heuristic, as its shaping rate can be adapted by changing the threshold weight level $\ell_\text{max}$.

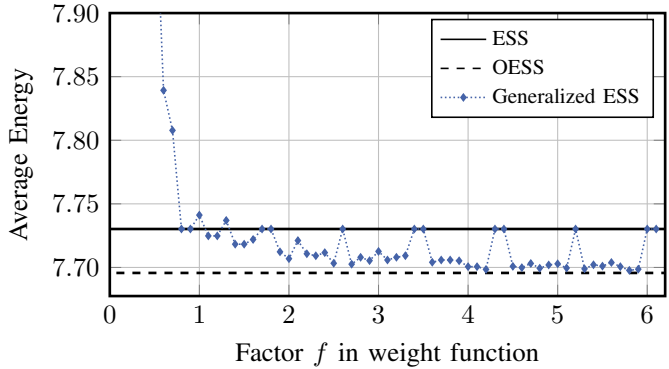
\begin{figure}
  \begin{tikzpicture}
    \begin{axis}[
        xlabel={Factor $f$ in weight function},
        ylabel={Average Energy},
        ymax={7.9},
        xmin={0},
        xmax={6.2},
        grid=both,
        width=1\columnwidth,
        height=0.6\columnwidth,
        line width=1pt,
        legend style={
          nodes={transform shape},
          font=\footnotesize,
          line width=0.6pt
        },
        legend cell align={left},
        y tick label style={
          /pgf/number format/.cd,
          fixed,
          fixed zerofill,
          precision=2,
          /tikz/.cd
        },%
      ]%
      \addplot[
      ]%
        coordinates {(-1, 7.730231285095215) (7, 7.730231285095215)};
      \addplot[
        dashed
      ]%
        coordinates {(-1, 7.695684432983398) (7, 7.695684432983398)};
      \addplot[%
          KITblue,
          mark=diamond*,
          mark options={solid},
          mark size=1.2pt,
          line width=0.6pt,
          densely dotted
        ]%
        coordinates {
          (0.1, 11.661834716796875)
          (0.2, 8.680301666259766)
          (0.30000000000000004, 8.194931030273438)
          (0.4, 8.221280097961426)
          (0.5, 8.01620864868164)
          (0.6, 7.83926248550415)
          (0.7000000000000001, 7.80776309967041)
          (0.8, 7.730237007141113)
          (0.9, 7.730237007141113)
          (1.0, 7.74116325378418)
          (1.1, 7.724809646606445)
          (1.2000000000000002, 7.724809646606445)
          (1.3000000000000003, 7.736970901489258)
          (1.4000000000000001, 7.718293190002441)
          (1.5000000000000002, 7.718293190002441)
          (1.6, 7.721991539001465)
          (1.7000000000000002, 7.730237007141113)
          (1.8000000000000003, 7.730237007141113)
          (1.9000000000000001, 7.712282180786133)
          (2.0, 7.70688533782959)
          (2.1, 7.721147537231445)
          (2.2, 7.710750102996826)
          (2.3000000000000003, 7.709187030792236)
          (2.4000000000000004, 7.711658954620361)
          (2.5000000000000004, 7.703284740447998)
          (2.6, 7.730237007141113)
          (2.7, 7.702572345733643)
          (2.8000000000000003, 7.707973003387451)
          (2.9000000000000004, 7.705357551574707)
          (3.0000000000000004, 7.712618350982666)
          (3.1, 7.705855846405029)
          (3.2, 7.707993507385254)
          (3.3000000000000003, 7.7092390060424805)
          (3.4000000000000004, 7.730237007141113)
          (3.5000000000000004, 7.730237007141113)
          (3.6, 7.704079627990723)
          (3.7, 7.705853462219238)
          (3.8000000000000003, 7.705853462219238)
          (3.9000000000000004, 7.705251693725586)
          (4.0, 7.700621128082275)
          (4.1, 7.700621128082275)
          (4.2, 7.698451042175293)
          (4.3, 7.730237007141113)
          (4.3999999999999995, 7.730237007141113)
          (4.5, 7.700858116149902)
          (4.6, 7.699843406677246)
          (4.7, 7.703137397766113)
          (4.8, 7.699404239654541)
          (4.9, 7.701972961425781)
          (5.0, 7.702881813049316)
          (5.1, 7.6995625495910645)
          (5.2, 7.730237007141113)
          (5.3, 7.698933124542236)
          (5.4, 7.70211124420166)
          (5.5, 7.701043128967285)
          (5.6, 7.703900337219238)
          (5.7, 7.700634479522705)
          (5.8, 7.6977362632751465)
          (5.9, 7.698652267456055)
          (6.0, 7.730237007141113)
          (6.1, 7.730237007141113)};%
      \legend{%
        ESS,%
        OESS,%
        Generalized ESS%
      };
    \end{axis}%
  \end{tikzpicture}%
  \caption{Average amplitude energy for a sequence length of $224$ symbols and a shaping rate of $1.5$ bit / amplitude. Generalized \gls{ess} uses a weight function with factor $f$ computed from a Maxwell-Boltzmann distribution with $\lambda = 0.1$ using~\eqref{eq:weight-function}.}%
  \label{fig::generalized-avg-energy}%
\end{figure}

While comparing generalized \gls{ess} to normal \gls{ess}, we noticed that generalized \gls{ess} can outperform \gls{ess} by a small margin, even on the \gls{awgn} channel.
If only a part of the sequences represented by the \gls{ess} trellis are used for transmission, \gls{ess} suffers from a small rate loss.
Not using all sequences is often caused by transmitting a fixed number of bits while the number of sequences in the trellis is not a power of two.
Depending on the factor $f$ used in the weight function~\eqref{eq:weight-function}, the number of sequences in the generalized \gls{ess} trellis changes.
First observations suggest that this often leads to the number of unused sequences being smaller compared to \gls{ess}.
This in turn leads to a smaller rate loss and thus a reduced average energy.
Fig.~\ref{fig::generalized-avg-energy} demonstrates how varying $f$ influences the average energy of the code book, which directly relates to the rate loss.
Evidently, the average energy of the generalized \gls{ess} code book is less than the average energy of the \gls{ess} code book for most plotted values of $f$.
If $f \geq 4$, the average energy of the generalized \gls{ess} code book is frequently very close to the lower bound provided by the average energy of an \gls{oess}~\cite{chen_optimization-analysis_2022} code book.

\section{Conclusion} \label{sec:conclusion}

In this paper, we generalized \gls{ess} to allow arbitrary distributions and showed how even multiple amplitudes with the same probability may be handled.
The proposed generalization was used to achieve rate adaption for different channel qualities on a \gls{ppc} channel.

Future research could further analyze the observation that generalized \gls{ess} can reduce the rate loss of \gls{ess}.
Achieving this using a small scaling factor $f$ could yield a method to reduce the \gls{ess} rate loss with reasonable additional complexity.

\appendices

\section*{Acknowledgment}
This work has received funding from the European Research Council (ERC)
under the European Union's Horizon 2020 research and innovation programme (grant agreement No. 101001899) and the Deutsche Forschungsgemeinschaft (DFG, German Research Foundation) -- Grant 555885380.

\ifCLASSOPTIONcaptionsoff
  \newpage
\fi

\end{document}